          [2004/06/22 1.2 (KOF); 2001/04/25 1.1 (PWD)]

\documentclass[a4paper,twocolumn]{Gaia2004} 
\usepackage{times}      
\usepackage{epsfig}     
\usepackage{natbib}     

\title{Structure of the Galactic Halo towards the North Galactic Pole}

\author{T.D. Kinman}
\affil{NOAO,P.O.\ Box 26732, Tucson, AZ 85726-6732 - USA\footnote{\MakeUppercase{Noao} is operated 
by \MakeUppercase{Aura} \MakeUppercase{I}nc.\ under contract with the 
\MakeUppercase{N}ational \MakeUppercase{S}cience \MakeUppercase{F}oundation.}}
\author{A. Bragaglia} 
\author{C. Cacciari} 
\author{A. Buzzoni}
\affil{INAF, Osservatorio Astronomico Bologna, Via Ranzani 1 40127 Bologna - Italy}
\author{A. Spagna}
\affil{INAF, Osservatorio Astronomico Torino, Via Osservatorio 20 10025 Pino Torinese - Italy}

\bibpunct{(}{)}{;}{a}{}{,}  

\begin{document}


\maketitle

\begin{abstract}
\vskip -4mm

We have used RR Lyrae and Blue HB stars as tracers of the old Galactic halo, 
in order to study the halo structure and the galactic rotation as a function 
of height above the plane. Our sample includes 40 RR Lyrae and 80 BHB stars 
that are about 2 to 15 kpc above the plane, in a roughly 250~deg$^2$ area  
around the North Galactic Pole (NGP). We use proper motions (derived from the GSC-II
database) and radial velocities to determine the rotation of the halo. 
From the whole sample the motion appears to be significantly more retrograde 
than the samples in the solar neighborhood, confirming \citet{maj92} 
results and our own preliminary results based on 1/3 the present sample 
\citep[][]{kin03,spa03}. 
However, the better statistics has now revealed the likely existence of two 
components, whose characteristics need an accurate analysis of systematic 
errors on the proper motions in order to be assessed in detail.  
\end{abstract}

\section{Introduction}
\vskip -4mm

The kinematics of halo stars in the solar neighborhood ($Z \le 2$~kpc) is 
well established: mean heliocentric rotation velocity $\langle V \rangle \sim -220 \pm 10$~km\,s$^{-1}$ 
(from RR Lyrae and K giant stars, cf.\ \citealp{mar98,lay96,chi98,dam01}).\\
From halo stars observed at larger distances ($Z \ge 5$~kpc), the kinematics 
of the high halo is less clear: the mean rotation was found retrograde  
($\langle V\rangle = -275 \pm 16$~km\,s$^{-1}$) by \citet{maj92} and \citet{maj96} in SA 
57 near the NGP, and confirmed by \citet{gil02} in other directions.\\ 
However, other estimates calculated from orbital parameters \citep[][]{car99,chi00}, 
as well as \citet{sir04} recent analysis of very 
distant BHB stars from the SDSS, do not support any significant halo rotation. 
The present work attempts to resolve this discrepancy; it grew from earlier studies
of halo stars in the North Galactic Cap \citep[][]{kin96} which confirmed the 
streaming motion (in the W vector) that Majewski found for his subdwarf sample 
in SA 57.\\
So the questions 
are: {\it i) Is the high halo in retrograde rotation?
ii) Is there any halo substructure that may mimic  retrograde rotation in some
directions?}

\section{Our Data}
\vskip -4mm

\begin{figure}[!t]
\centerline{\epsfig{file=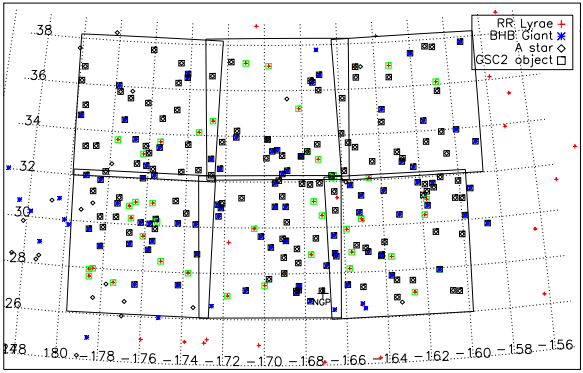,angle=90,width=\hsize,clip=}}
\vskip -3mm
\caption{The POSS fields at the NGP where we have performed the present analysis. 
Right ascension, along the X axis, is expressed in arc degrees.}
\label{fig:fields}
\end{figure}

Our sample of halo tracers consists of 80 blue HB (BHB) and 40 RR Lyrae (RRL) 
stars spread over an area of approximately $22^\circ\times12^\circ$ around the NGP,
at a distance $Z \simeq 1.5$ to $\sim 16$~kpc above the galactic plane. This area 
is the combination of 6 POSS fields where proper motions from the GSC-II 
database were measured (see later), and is shown in Fig.~\ref{fig:fields}.  
We refer the reader to \citet{kin03} for details on the target selection. \\
For these stars we have:\\
\vskip -5mm
$\bullet$ Absolute Magnitudes ($\pm$0.2 mag) $\Longrightarrow$ distances. 
In particular, $E(B-V)$ values are from \citet{schle98};
for the BHB stars, $M_V$ have been estimated from the $M_V$ vs.\ $(B-V)$ relation 
given by \citet{pre91} adjusted so that $M_V = 0.6$ at $(B-V) = 0.2$; 
for the RRab stars, we used $M_V = 0.22\, [Fe/H] + 0.93$ if $[Fe/H]$ was available, or 
$M_V = -1.619\, \log P + 0.20$ \citep{kin02} if only the period was available, or else 
$M_V = 0.6$. For the RRc stars it is assumed that $M_V = 0.6$.\\
\vskip -5mm
$\bullet$ Proper Motions (formal r.m.s. errors $\le 3$~mas\,yr$^{-1}$).  
These are based on the plate material used for the construction of the GSC-II 
catalogue using the method described in \citet{spa96}.\\
\vskip -5mm
$\bullet$ Radial velocities ($\pm 10-50$~km\,s$^{-1}$), obtained at the 4m-RC (KPNO) 
and 3.5m-LRS (TNG).\\
\vskip -5mm
From the above data we have derived heliocentric space velocity components U, 
V, and W using the program by \citet[updated version for the J2000 reference 
frame]{john87}, with a further update of the transformation 
matrix derived from the Volume 1 of the Hipparcos data catalogue. Finally, 
the heliocentric UVW velocities have been corrected adopting the solar motion 
${\rm (U,V,W)_\odot=(10.0,\,5.25,\, 7.17)}$~km\,s$^{-1}$ with respect to the LSR,
from  \citet{db98}.

\section{Errors}
\vskip -4mm

\begin{figure}[!t]
\centerline{\epsfig{file=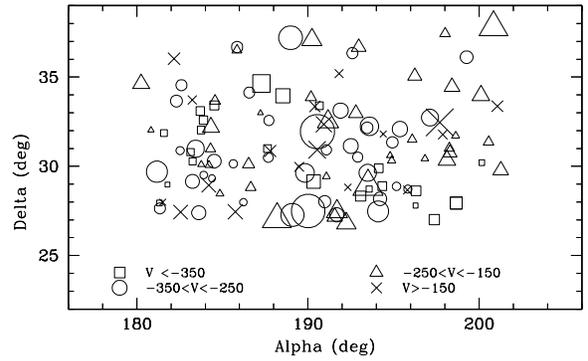,width=\hsize,clip=}}
\vskip -3mm
\caption{Rotational velocity V after correction for pm systematic errors 
in each field, as a function of position and distance. Different symbols 
represent different V intervals, and the size of the symbol is proportional 
to the parallax.} 
\label{fig:nobias}
\end{figure}

An error in $M_V$ will not have a large effect on the derived U and V 
velocities (a total uncertainty of as much as $\pm 0.2$~mag would
correspond to $\pm 10$\% in the distance and hence V). The U and V velocities
are nearly independent of the radial velocity in the direction of the NGP.\\
{\em In this case, the greatest uncertainty in V will come from errors in the 
proper motions.}\\
Therefore it is important to test the systematic errors in the proper motion 
(pm) data using the QSOs and galaxies present in the observed fields. 
Proper motions have been derived for 65 QSOs and 50 galaxies, omitting a few 
objects whose pm or errors exceed 10~mas\,yr$^{-1}$  \citep{kin03}. 
These 115 objects, which should have zero proper motion, have the 
following average GSC-II pm:\\ 
\begin{equation}
\left\{
\begin{array}{lll}
\mu_\alpha &=& -0.109 \pm 0.122~{\rm mas\, yr}^{-1}\\
\mu_\delta &=& -0.422 \pm 0.162~{\rm mas\,yr}^{-1}
\end{array}
\right.
\end{equation}
Presumably these systematic errors ($< 1$~mas\,yr$^{-1}$, comparable to those of the 
pm measured by Hipparcos for much brighter stars) apply to our program stars 
over the same sky area and a similar magnitude range.\\  
A preliminary check shows systematic errors of similar size in our individual 
fields.  Although these individual field errors are necessarily
more poorly determined, we have used them to correct our BHB and RRL 
motions in this preliminary analysis. The distribution of V shows little systematic 
trend with either position or distance (see Fig. \ref{fig:nobias}), 
so these corrections introduce little or no bias.\\
Currently, the accuracy of the GSC-II proper motions represents the best 
available for the large-field surveys that are needed for studies of
Galactic structure; higher precision is necessary but must await dedicated 
space missions such as GAIA.

\begin{table*}[!t]
\caption{Space motion vectors for the present halo stellar sample, and for a local halo sample 
for comparison. }
\label{tab:uvw}
\begin{center}
\leavevmode
\begin{tabular}[h]{cccccccl} 
\noalign{\smallskip}
\hline 
\noalign{\smallskip}
  U    &   V   &   W  &$\sigma_{\rm u}$&$\sigma_{\rm v}$&$\sigma_{\rm w}$& No. of stars& Source \\
\noalign{\smallskip}
\hline 
\noalign{\smallskip}
$-$12$\pm$15 &$-$256$\pm$9 & $-$10$\pm$9 & 153 & 92 & 91 & 108 (101 for W) &  Present sample \\
$+$4$\pm$15 &$-$264$\pm$9 & $-$20$\pm$9 & 158 &  91 & 95 & 82 (75 for W) &  Present sample at Z $>$ 4 kpc \\
$-$1$\pm$26 &$-$219$\pm$24 & $-$5$\pm$10 & 193 & 91 & 96 & 84  & Martin \& Morrison (1998) local sample   \\
\noalign{\smallskip}
\hline 
\end{tabular}\end{center}\end{table*} 

\section{Results}
\vskip -4mm

\begin{figure}[!ht]
\centerline{\epsfig{file=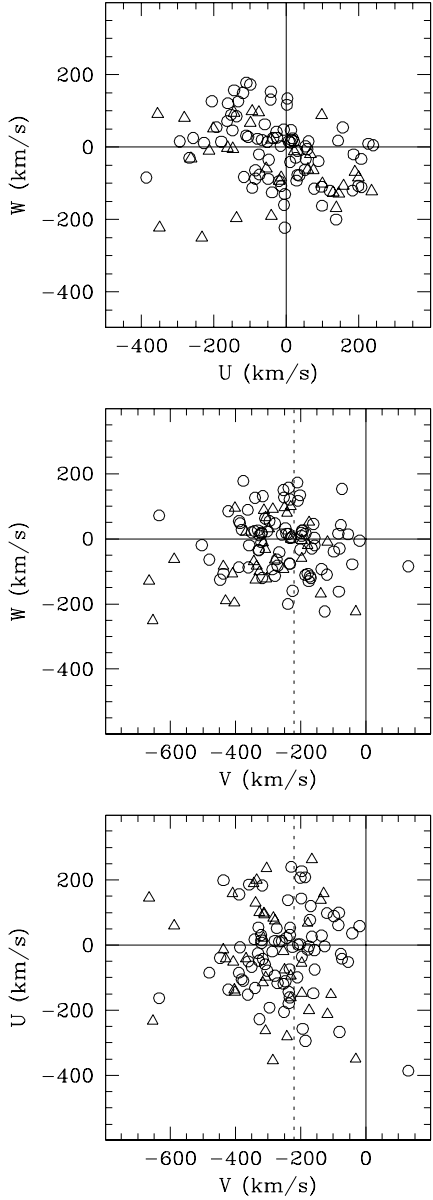,width=0.78\hsize,clip=}}
\vskip -3mm
\caption{Plots W--U, W--V and U--V of the entire stellar sample, including 
the ``outliers'' that were trimmed out in the previous considerations. 
Triangles and circles represent RR Lyrae and BHB stars, respectively. 
The mean heliocentric rotation velocity $\langle V \rangle \sim -220 \pm 10$~km\,s$^{-1}$ is reported
in the V plots (shaded line). Note that in the different velocity planes the contamination 
from disk stars is always quite small.}
\label{fig:uvw}
\end{figure}

In calculating the UVW vectors we put the radial velocity equal to zero (with 
an error of $\pm 150$~km\,s$^{-1}$) if no radial velocity was available. In such cases
(7 RRL stars at $Z > 4$~kpc), the U and V vectors should be scarcely affected 
but the W vector is discarded.
For a better characterization of our stars, we have also trimmed 10\% of the 
most deviating stars from our sample when estimating mean values or distributions. 
We have found that trimming has little effect on the mean values of U, V and W, 
although it does reduce the velocity dispersions $\sigma_{\rm U}$, $\sigma_{\rm V}$ 
and $\sigma_{\rm W}$. 
In Table~\ref{tab:uvw} we compare the mean heliocentric UVW values of our entire 
sample and of the subsample at $Z > 4$~kpc with those found by \citet{mar98}
for their {\sc Halo2} sample of local RRL stars (in all cases 10\% 
of outliers were trimmed). We also show the W--U, W--V and U--V plots 
in Fig.~\ref{fig:uvw}.\\
A KMM test \citep[cf.][]{ash94} on the entire trimmed sample gives 
{\em about 90\% probability that the sample is formed by two groups, one containing 
$\sim 56$\% of the stars with estimated $\langle V \rangle = -316 \pm 8$~km\,s$^{-1}$, and the other 
containing $\sim 44$\% of the stars with estimated $\langle V \rangle = -177 \pm 9$~km\,s$^{-1}$, 
same dispersion $\sigma = 61$~km\,s$^{-1}$.} 
We show in Fig.~\ref{fig:hist} the distributions of V as a function of the 
distance $Z$ above the Galactic plane, where the bimodal shape is evident at all 
distance intervals as well as for the entire sample.

\section{Conclusions}
\vskip -4mm

\begin{figure}[!t]
\centerline{\epsfig{file=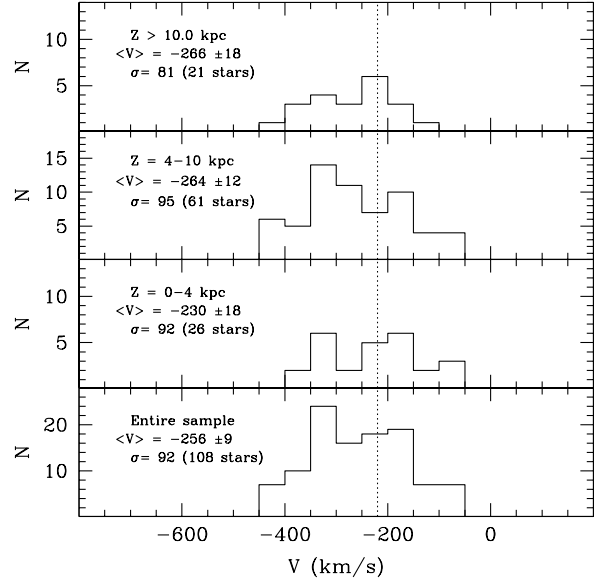,width=\hsize,clip=}}
\caption{Histograms of rotational velocity V as a function of distance 
Z above the Galactic plane: all distributions suggest a bimodal structure. }
\label{fig:hist}
\end{figure}

Our present results, based on a 3 times larger sample than our previous
preliminary  analysis \citep[][]{kin03,spa03}, seem to lead to
the following conclusions:\\
\vskip -5mm
$\bullet$  For $Z < 4$~kpc, the mean rotation V is close to that found for the halo in the 
solar neighborhood (see corresponding panel in Fig.~\ref{fig:hist}).\\
\vskip -5mm
$\bullet$ For $Z > 4$~kpc, the {\em mean rotation is significantly more retrograde} - in 
agreement with Majewski's earlier finding.  
This seems to be due to a {\em group of very retrograde stars}, 
whose relative importance  increases with $Z$ with respect to the dissipative halo
component, and is particularly significant  in the range $Z = 4-10$~kpc (see 
Fig.~\ref{fig:hist}). Therefore, this is likely to be a {\em local substructure, 
maybe associated with an accretion event}, since  \citet{sir04} analysis
of a large number of distant BHB stars spread over the sky  shows little
deviation from the solar neighborhood (dissipative halo) value.\\ 
\vskip -5mm
It will be interesting to extend the work to halo stars in Anticentre fields. 
Possibly this will allow us to detect gradients in the V motion and discover
whether this  is a local effect limited to the NGP or part of a larger systematic 
effect.\\
\vskip -5mm
The errors in the GSC-II are small enough to make our results worth consideration. 
On the other hand, they are large enough to make it clear that 
{\em  GAIA proper motions are essential for a detailed
knowledge of the Galactic structure}.

\section*{Acknowledgments}
\vskip -4mm

The GSC-II is a joint project of the Space Telescope Science
Institute (STScI) and the INAF-Osservatorio Astronomico di Torino
(INAF-OATo). We acknowledge the GSC-II team and in particular M.G.
Lattanzi and R.L. Smart for their valuable support to this study.\\
The present work has been supported by the MIUR (Mi\-ni\-ste\-ro dell'Istruzione, 
dell'Universit\`a e della Ricerca).

\vfill\eject


\begin{thebibliography}{}

\bibitem[Ashman, Bird \& Zepf(1994)]{ash94}  
	Ashman, K. M.,  Bird, C. M., Zepf S. E. 1994, AJ, 108, 2348

\bibitem[Carney(1999)]{car99}  
	Carney, B. W. 1999, in {\it The Third Stromlo Symposium: 
	The Galactic Halo}, eds. Gibson, B. K., Axelrod, T. S.,  Putman, M. E., ASP 
	Conf.\ Ser.\ Vol.\ 165, p.\ 230

\bibitem[Chiba \& Yoshii(1998)]{chi98}  
	Chiba, M.,  Yoshii, Y. 1998, AJ, 115, 168 

\bibitem[Chiba \& Beers(2000)]{chi00}  
	Chiba, M.,  Beers, T. C. 2000, AJ, 119, 2843

\bibitem[Dambis \& Rastorguev(2001)]{dam01}  
	Dambis, A. K.,  Rastorguev, A. S. 2001,  Pis'ma Astron. Zh. 27, 132 

\bibitem[Dehnen \& Binney(1998)]{db98} 
	Dehnen, W.,  Binney, J. J. 1998, MNRAS, 298, 387

\bibitem[Gilmore, Wyse \& Norris(2002)]{gil02} 
	Gilmore, G., Wyse, R. F. G.,  Norris, J. E. 2002, ApJ, 574, L39

\bibitem[Johnson \& Soderblom(1987)]{john87}  
	Johnson, D. R. H,,  Soderblom, D. R. 1987,  AJ, 93, 864

\bibitem[Kinman(2002)]{kin02}  
	Kinman, T. D. 2002, IBVS 5354

\bibitem[Kinman et al.(1996)]{kin96}  
	Kinman, T.D., Pier, J. R., Suntzeff, N. B., {\it et al.} 1996, AJ, 111, 1164 

\bibitem[Kinman et al.(2003)]{kin03}  
	Kinman, T. D.,  Cacciari, C., Bragaglia,  A., Buzzoni, A., Spagna, A. 2003, 
	in {\it Galactic,  Stellar Dynamics}, Proc.\ of JENAM 2002, eds.\ C. M. Boily, 
	P. Patsis, S. Portegies-Zwart, R. Spurzem,  C. Theis, EAS Pub. Ser.,  Vol.\ 10, p.\ 115

\bibitem[Layden et al.(1996)]{lay96}  
	Layden, A. C.,  Hanson, R. B.,  Hawley, S. L.,  Klemola, A. R., Hanley, C. J. 1996, AJ, 112, 2110

\bibitem[Majewski(1992)]{maj92}  
	Majewski, S. 1992, ApJS, 78, 87 

\bibitem[Majewski, Munn \& Hawley(1996)]{maj96}  
	Majewski, S.,  Munn, J. A., Hawley, S. L. 1996, ApJ, 459, L73

\bibitem[Martin \& Morrison(1998)]{mar98}  
	Martin, J. C., Morrison, H. L. 1998, AJ, 116, 1724

\bibitem[Preston, Shectman \& Beers(1991)]{pre91}  
	Preston, G. W., Shectman, S. A., Beers, T. C. 1991, ApJ, 375, 121

\bibitem[Schlegel, Finkbeiner \& Davies(1998)]{schle98}  
	Schlegel, D. J.,  Finkbeiner, D. P., Davis, M. 1998, ApJ, 500, 525

\bibitem[Spagna et al.(1996)]{spa96} 
	Spagna, A., Lattanzi, M. G., Lasker, B. M., McLean, B. J., Massone, G.,  Lanteri, L. 1996, 
	A\&A, 311, 758     

\bibitem[Spagna et al.(2003)]{spa03}  
	Spagna, A.,  Cacciari, C.,  Drimmel, R.,  Kinman, T.D., Lattanzi, M.G., Smart, R.L. 
	2003, in {\it Gaia  Spectroscopy, Science,  Technology}, ed.\ U.Munari, ASP Conf. Ser. 
	Vol.\ 298, p.\ 137 

\bibitem[Sirko et al.(2004)]{sir04}  
	Sirko, E., Goodman, J., Knapp, G. R., {\it et al.} 2004, AJ, 127, 914
 \end{thebibliography}
\end{document}